# Emulation of Neuron and Synaptic Functions in Spin-Orbit Torque Domain Wall Devices

Durgesh Kumar[1], Ramu Maddu[1], Hong Jing Chung[2], Hasibur Rahaman[1], Tianli Jin[1], Sabpreet Bhatti[1], Sze Ter Lim[2], Rachid Sbiaa[3], S. N. Piramanayagam[1,*]

[1]School of Physical and Mathematical Sciences, Nanyang Technological University, 21 Nanyang Link, Singapore, 637371

[2]Institute of Materials Research and Engineering, A*STAR, 2 Fusionopolis Way, Innovis, Singapore, 138634

[3]Department of Physics, Sultan Qaboos University, Muscat, Oman

[*]*prem@ntu.edu.sg*

## Abstract

Neuromorphic computing (NC) architecture has shown its suitability for energy-efficient computation. Amongst several systems, spin-orbit torque (SOT) based domain wall (DW) devices are one of the most energy-efficient contenders for NC. To realize spin-based NC architecture, the computing elements such as synthetic neurons and synapses need to be developed. However, there are very few experimental investigations on DW neurons and synapses. The present study demonstrates the energy-efficient operations of neurons and synapses by using novel reading and writing strategies. We have used a W/CoFeB-based energy-efficient SOT mechanism to drive the DWs at low current densities. We have used the concept of meander devices for achieving synaptic functions. By doing this, we have achieved 9 different resistive states in experiments. We have experimentally demonstrated the functional spike and step neurons. Additionally, we have engineered the anomalous Hall bars by incorporating several pairs, in comparison to conventional Hall crosses, to increase the sensitivity as well as signal-to-noise ratio (SNR). We performed micromagnetic simulations and transport measurements to demonstrate the above-mentioned functionalities.

# Introduction

Artificial intelligence (AI) finds growing interest in day-to-day consumer devices such as smartphones, tablets, laptops, televisions, self-driving cars, and many more [1]. AI potentially performs popular cognitive tasks like image/fingerprint recognition, big-data analysis, and unmanned vehicle control in the above-mentioned consumer devices [1]. Currently, these intelligent functions are being performed on conventional computing architectures, based on von Neumann formalism, by using machine learning algorithms. In von Neumann's architecture, a huge amount of data travels back and forth between the central processing unit (CPU) and the memory unit [2]. This limits the speed of the computing system, which is known as the "von Neumann bottleneck". The speed problem inflames further due to the difference in the operating speeds of the faster CPU and relatively slower memory unit, called the "memory wall". Above all, the present computing system is energy inefficient. For instance, a huge amount of power (~of the order of MW) is required to simulate the activities of a cat's brain in the present computing architecture. In contrast, the human brain requires only a few watts of power to compute the same problem. Therefore, the research community is looking for alternative computing architectures that can mimic the human brain at a low energy cost. Neuromorphic computing (NC), also known as brain-inspired computing, is one such computing architecture [2]. The brain comprises billions of neurons, which are interconnected through trillions of synapses. Here, neuron act as processing unit, which receives multiple inputs from pre-synaptic neurons and generates an output. The synapse, however, acts as a memory unit. The synapses also form the bridge between pre-synaptic neurons and post-synaptic neurons and control the weightage of information flowing between them.

In the past, several types of devices have been investigated to realize NC. These devices include resistive memristors, phase change random access memory (PCRAM), and ferroelectric materials-based RAM (FeRAM) [3-6]. Besides these candidates, spintronic devices - which were mostly studied for memory devices - have also shown the potential to be used as synthetic neurons and synapses. Spintronic devices promise ultra-low energy and high endurance. These spintronic devices include spin-torque nano oscillators (STNO), superparamagnetic magnetic tunnel junctions (MTJs), spin Hall nano-oscillators (SHNOs), domain wall (DW) devices, magnetic skyrmions, and antiferromagnetic materials-based heterostructures [7, 8]. Amongst these spintronic devices, DW devices are one of the most energy-efficient candidates that can readily be implemented [9]. Moreover, the field of spin-orbit torque (SOT), which is used to drive the DWs, is only at the research development stage. Therefore, there is a lot of scope for the energy requirement aspect.

The synaptic functions require multiple resistance states in DW devices. This can be achieved by controllably stopping the DWs at multiple desired positions. For instance, Borders et al. fabricated Hall Bar devices from $(Co/Ni)_n$ ferromagnetic (FM) wire and demonstrated multiple resistance states [10]. The Laplace force on the DW at the Hall bar junction (the parking site for DWs, known as the pinning site) generates effective pinning. Later, Cai et al. fabricated a DW device (from a full MTJ stack) with increasing width of the wire and demonstrated the multiple-resistance states [11]. Similar to the previous case, the Laplace force on the DWs was utilized to pin the DWs. The Laplace force may not be able to generate strong pinning on the DWs, therefore, deeper research may be needed to improve the suitability of these methods for applications. Jin et al. studied the DW motion in an FM wire with local ion implantation and demonstrated the multiple-resistance states [12]. Siddiqui et al. demonstrated the multi-resistance states in the MTJ device with a discrete reference layer [13]. More recently, Kumar et al. fabricated the pinning sites by altering the interfacial Dzyaloshinskii-Moriya interaction (*i*DMI) locally and showed multi-resistance states using micromagnetic simulations [14]. Similarly, Liu et al. achieved synaptic functions in the DW devices with double-sided notches [15].

In NC, neurons receive inputs from several other neurons and once the strength of the input(s) exceeds a threshold value, an output is generated. In addition, the neurons need to be reset just after generating an output. Hasan et al. performed simulations and studied a neuron device where a stray field from a neighboring FM wire causes the leaking [16]. Later, Mah et al. proposed and studied a neuron device with a graded anisotropy field ($H_k = 2K_u/M_s$) [17]. Sato et al. recorded GHz oscillations of DW in the z-shaped DW devices for fabricating artificial neurons [18]. Moreover, Kumar et al. performed micromagnetic simulations in an FM wire with local *i*DMI and observed that the DW oscillates at GHz frequency [19]. A few other researchers studied the oscillations of the DWs in the DW devices for fabricating synthetic neurons [20, 21].

Bhowmik et al. studied a neural network where synapses were designed using the DW devices and neurons were designed using transistors [22]. Sharad et al. developed an energy-efficient neural network based on DW devices, which consumed ~100× less energy compared to complementary metal-oxide semiconductor (CMOS)-based architecture, to perform image recognition tasks [23]. Later, the same group developed a neural network that was 250× energy efficient compared to CMOS architecture [24]. Several other studies reported the usefulness of DW devices for neuromorphic architecture [25, 26].

At present, most of the studies in the field of DW-based neuromorphic computing are simulation-based and experimental work is only at the primitive stage. In this paper, we have studied DW device-based synapses and neurons using micromagnetic simulations as well as experiments. For synaptic applications, we proposed the idea of meander devices, where two segments of the DW devices join at an offset of "*d*" (figure 1 (a)). When the DW reaches the pinning site, it experiences different energy. In other words, the DW needs higher energy to move further from this position. Moreover, the strength of the pinning increases as *d* increases. In simulations, we studied a large range of "*d*" starting from 10% to 90%, in steps of 10%. Based on our theoretical findings, we fabricated DW devices with *d* = 40%, 50%, and 60% in the experiments. We utilized SOT from the W layer to move the DWs at low current densities. We employed a novel reading strategy to get the output in the form of the electrical voltage. This novel design helped us to improve the sensitivity and enhance the signal-to-noise ratio (SNR) of the output signal.

For neuron devices, we studied the DW motion in straight DW devices. Similar to the previous case, we first performed micromagnetic simulations to understand the underlying physics of SOT-driven DW motion in neuron devices. Subsequently, we fabricated neuron devices and performed the experimental characterization. For the case of spike neurons, we fabricated conventional Hall bar electrodes at the right end of the device. Once the DW reaches the vicinity of the read probe, a spike in the output voltage occurs. However, step neurons utilize the novel reading scheme. The details of our simulation and experimental results will be discussed in the relevant sections of the paper.

# Results and Discussions
## Domain Wall-Based Synaptic Devices

### 1. Micromagnetic Simulations

To systematically understand the SOT-driven DW dynamics in DW-based synaptic and neuron devices, we first performed micromagnetic simulations using Mumax$^3$ [27]. At first, we studied two types of DW devices, (*i*) meander DW devices (synapse) and (*ii*) straight DW devices (neuron). The dimensions of the above devices were taken as 256 nm × 32 nm × 1 nm. The cell size was taken as 1 nm × 1 nm × 1 nm. Based on the magnetic parameters that we used in our simulations, the exchange length is ~5 nm, which is much larger than the cell size in all the axes [28]. The details of the micromagnetic parameters are presented in the methods section.

In all the simulations, we first inserted a ↓→↑ type DW, which is supported by the given *i*DMI constant of the material system [9]. Subsequently, we applied the electrical current (DC/pulsed current) to push the DW from one end of the device to the other. The magnitude of the current density (*J*) was varied from $1 \times 10^{10}$ A/m$^2$ to $1 \times 10^{12}$ A/m$^2$. The *z*-component of the magnetization ($m_z$) was utilized to represent the DW position in our simulations.

First, we simulated the DW motion in meander-shaped DW devices with one pinning site (two meander segments). As mentioned earlier, we varied "*d*" from 10% to 90% (in the steps of 10%) in simulations to get a deeper insight. As shown in figure 1 (b), when a *J* of $5 \times 10^{10}$ A/m$^2$ is applied, the DW moves from the left end of the wire and gets pinned at the pinning site for "*d*" of 30% and higher. The DW does not pin for the offset values of 10% and 20%. The inset of figure 1 (b) shows the stable magnetization state after the DW gets pinned for the case of *d* = 40% and *J* = $5 \times 10^{10}$ A/m$^2$. A similar stable magnetization state was observed in all the other pinning events.

Subsequently, we performed these simulations for the full range of current densities. For an offset of 10%, the DW does not pin at any of the studied current densities. However, the DW pinning was observed only for *J* = $1 \times 10^{10}$ A/m$^2$ in the meander devices with *d* = 20%. For higher offset values, the DW pinning was observed for a large range of current densities. Notably, no domain wall depinning was observed for *d* = 90%. Please refer to supplementary section 1 for the corresponding phase diagram. We defined the current density beyond which the DW does not pin at the pinning site as depinning current density ($J_{dep}$) [29]. As shown in figure 1 (c), the depinning current density increases with the increase in the offset. This confirms our hypothesis that the pinning strength increases as "*d*" increases and it becomes extremely strong at *d* = 90%.

Based on the results for meander devices with one pinning site, we then studied the DW motion in the wire with 8 pinning sites (9 meander segments). The aspect ratio (*w*:*l*, figure 1 (a)) for every meander segment was kept at 1:4, the same as devices with one pinning site. Therefore, the total length of the devices increased to 1152 nm. All the other parameters were kept the same as in the previous simulations. The current density with a magnitude of $1 \times 10^{12}$ A/m$^2$ was utilized in the form of pulses. Besides, the pulse width was optimized in such a way that the DW travels from one pinning site to the next in one pulse. Similar to previous simulations, we first inserted ↓→↑ DW in the left-most segment, and then a current pulse ($|J| = 1 \times 10^{12}$ A/m$^2$) was applied to push the DW to the first pinning site. Subsequently, the current was switched off for 2 ns. During this period, the DW stays at the first pinning site. Then another current pulse was applied to push the DW to the second pinning site and again the current was switched off for 2 ns. This process continues until the DW sweeps the whole device. The results of this set of simulations are presented in figure 1 (d-e). Here, every image illustrates the instantaneous magnetization state at a given simulation time (as indicated in yellow color). The blue and red colors represent the magnetizations in -z and +z directions, respectively (inset of figure 1 (b)). We plotted the domain wall position ($m_z$) vs simulation time ($t$) (figure 1 (e)). We have observed systematic pinning and depinning of the DW at every pinning site, resulting in ten multiple resistance states. The inset of figure 1 (e) shows the profile of the pulsed current applied during these simulations. The DW exhibits damped oscillatory motion before it settles down at every pinning site. After understanding the detailed domain wall dynamics in simulations, we performed experiments to demonstrate the synaptic performance.

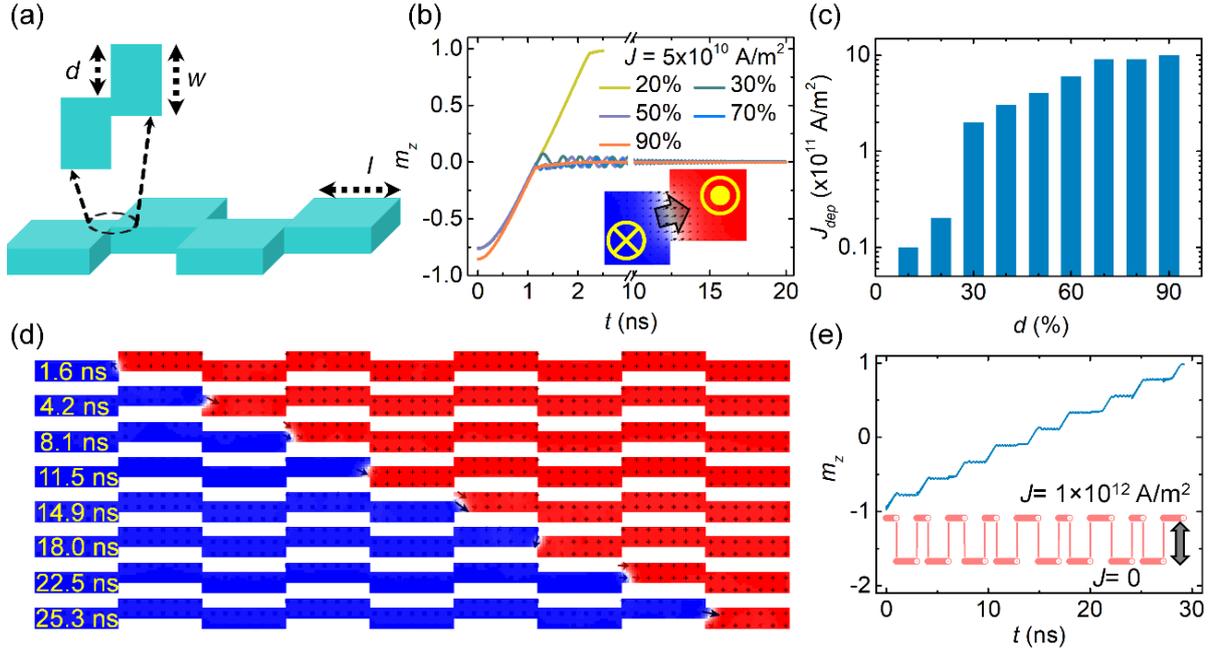

*Figure 1.* (a) Schematic diagram of the meander-shaped synaptic devices. Here, the domain wall (DW) device segments (with the length and widths of l and w, respectively) connect at an offset of "d". (b) The plot of "$m_z$" vs "t" for a meander device with one pinning site. These simulations were performed for different offset values ranging from 10 to 90% in the steps of 10%. All these simulations were performed at $J = 5 \times 10^{10}$ A/m$^2$. The figure in the inset shows the stable magnetization state when the DW is pinned at the pinning site. (c) The graph of $J_{dep}$ as a function of offset "d". (d) The instantaneous magnetization state at different simulation times to demonstrate the DW pinning at successive pinning sites. (e) The "$m_z$" vs "t" graph, which illustrates the intermediate magnetization states. The inset is the corresponding current density profile.

## 2. Experimental Results

We have deposited the film stack Si/SiO$_2$/ HP-W (3 nm)/ LP-W (3 nm)/ Co$_{40}$Fe$_{40}$B$_{20}$ (1 nm)/ MgO (1 nm)/ Ru (2 nm) [inset of figure 2 (a)] using DC/RF sputtering process. Here, HP and LP mean the high- and low-pressure deposition as can be seen in the methods section for more details on the thin film deposition parameters. Afterward, we measured the magnetic hysteresis (M-H) loop and observed a good perpendicular magnetic anisotropy (PMA) in our samples.

These results are presented in figure 2 (a). The presence of PMA in an as-deposited state is a hint of good SOT properties. However, the SOT properties were quantified using Harmonic measurements and the results are presented in reference [9]. The coercivity of the samples was found to be 44 Oe. Moreover, the evolution of Kerr images with increasing magnetic field suggests domain wall motion-mediated magnetization reversal in our samples.

To experimentally study the DW dynamics in our synaptic and neuron devices, we fabricated micro-meter-sized meander and straight wires, respectively. The dimensions of these devices were 900 μm × 20 μm. Based on our simulation results, the offsets (*d*) of meander devices were chosen to be 40, 50, and 60% (with respect to the width of the DW devices). To achieve different functional outputs, we proposed a novel engineering design for fabricating the Hall bars. In conventional Hall bar design, the output is measured using one pair of pins [30]. In this case, the major portion of the output signal is collected from the regions in the vicinity of the Hall crosses. Any change in magnetization far from the Hall probes can not be measured (or a very weak signal) using this method. In contrast, we proposed and studied the following novel design. For the synaptic device, we need to collect the signal from all over the device. Therefore, we fabricated the eight pairs of equispaced Hall bar pins. Moreover, the dimensions of Hall crosses were chosen in such a way that the maximum writing current flows through the current channel/DW devices. A similar design of Hall bars was utilized for the case of step neurons. This helps in improving the sensitivity of the output signal as well as enhancing SNR. The reading strategy for synaptic and step neuron devices is schematically presented in figure 2 (b). However, for the case of spike neurons, we fabricated one pair of Hall bars (conventional design) at the right end of the device. Once the DW reaches the vicinity of the Hall bar, the output in the form of the spike can be obtained.

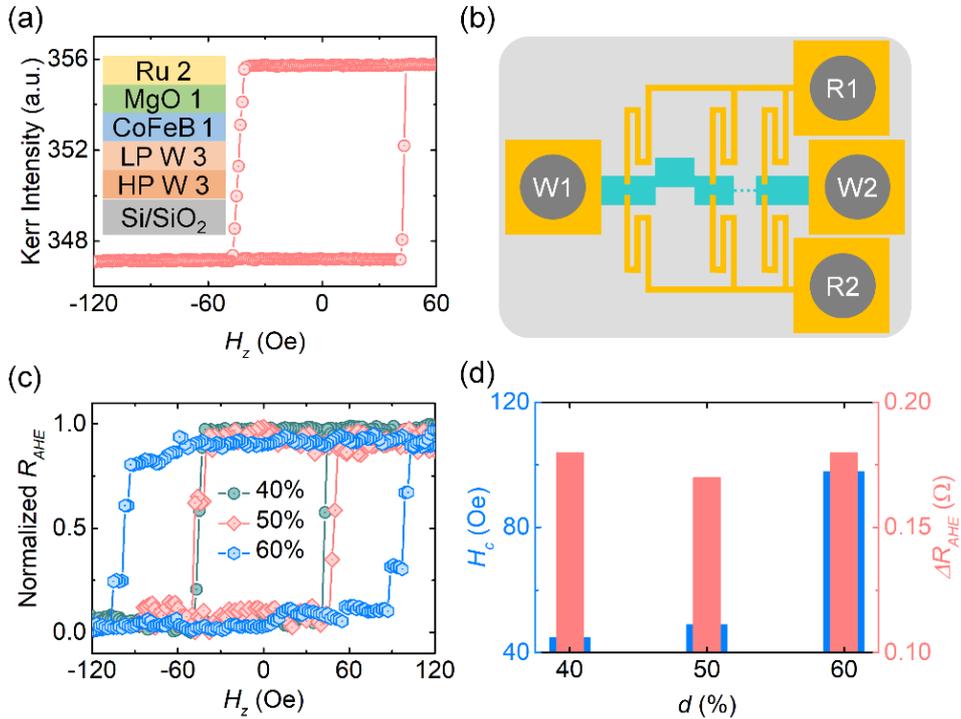

*Figure 2. (a) The magnetic hysteresis loop of the film stack used in this study. Inset: The schematic diagram of the film stack. (b) The schematic of the synaptic devices with uniquely designed read and write electrodes. (c) The R-H loop, illustrating the multiple resistance states in our synaptic devices. (d) The plots of $H_c$ and $\Delta R_{AHE}$ as a function of "d" in meander devices.*

Firstly, we measured the R-H loops of the synaptic devices, and the results are presented in figure 2 (c) (please refer to the methods section for the details of the experimental procedure employed in these experiments). For $d = 40\%$, we observed a sharp DW motion-mediated magnetization switching (without any intermediate magnetization state). However, once the offset increases to 50% & 60%, the intermediate magnetization states were observed. For instance, we observed a total of 3 states for $d = 50\%$ and 5 states for $d = 60\%$. This means that the pinning strength of the pinning sites increases as "*d*" increases. This result is consistent with the simulation results. Interestingly, the device coercivity ($H_{c\text{-}device}$) increases as "*d*" increases (figure 2 (d)). For $d = 40\%$, $H_{c\text{-}device}$ was found to be 45 Oe, which is almost equivalent to the thin film coercivity. However, the same increases to 98 Oe for $d = 60\%$, which suggests that the pinning strength increases as the offset value increases. In addition to the change in the coercivity, we also estimated the difference in anomalous Hall resistance between the up and down magnetization states and found similar values for all the meander devices (figure 2 (d)). This is expected as the read probe design is identical in all the synaptic devices. Additionally, we also performed the magnetization switching experiments while sweeping the electrical current (in the presence of an in-plane (IP) magnetic field) and observed the multi-resistance states. Please see supplementary section 2 for detailed discussions of the results from the current sweeping experiments. These results are also summarized in table I below.

*Table I:* *Summary of different properties of meander devices with different offsets (d).*

| $d$ (%) | $H_c$ (Oe) | $\Delta R_{AHE}$ (Ω) | $J_{dep}$ (A/m$^2$) | Number of Resistance States | | |
|---|---|---|---|---|---|---|
| | | | | $H_z$ Sweep | $I$ Sweep | Pulsed Current |
| 40 | 45 | 0.18 | $2.5 \times 10^{10}$ | 2 | 2 | 5 |
| 50 | 49 | 0.17 | $4 \times 10^{10}$ | 3 | 5 | 7 |
| 60 | 98 | 0.18 | $5 \times 10^{10}$ | 5 | 7 | 9 |

Next, we performed SOT-induced DW motion experiments in our synaptic devices. After saturating the devices, we applied the current pulses of a certain amplitude. The current density ($J$) was varied from $5 \times 10^9$ A/m$^2$ to $5 \times 10^{10}$ A/m$^2$ in most of the cases (unless specified). A simultaneous longitudinal magnetic field of 500 Oe was applied in all cases. The directions of the current, as well as the in-plane magnetic field, were kept the same as the current sweeping experiments and illustrated in figure 3 (a). As one can see in figure 3 (a) for the case of an offset of 40%, when the current pulses of $J = 3 \times 10^{10}$ A/m$^2$ together with an in-plane magnetic field of 500 Oe are applied in the synaptic devices, a reversed domain gets nucleated and the DW moves from the left end to the right end of the device. The nature of the DW motion is consistent with the SOT-driven DW motion in similar systems [9]. We repeated these experiments for synaptic devices with an offset of 50% and 60%. These results suggest that the pinning strength of the pinning sites is increasing as the magnitude of the offset increases. Again, these results are consistent with the previous results. Moreover, we also witnessed multiple resistive states as a result of uniform current pulses. These results are presented in detail in supplementary section 3.

To understand the DW dynamics in synaptic devices in greater detail, we plotted the phase diagram of DW dynamics as a function of offset and current density (figure 3 (b)). We have performed the experiments for the current density range starting from $5 \times 10^9$ A/m$^2$ to $5 \times 10^{10}$ A/m$^2$. However, we have shown the results from $1 \times 10^{10}$ A/m$^2$ to $5 \times 10^{10}$ A/m$^2$ for a clear presentation of the data. The results for $J = 5 \times 10^9$ A/m$^2$ are identical to that observed for $J = 1 \times 10^{10}$ A/m$^2$. Here, the violet, green, and yellow colors represent the "no domain wall motion",

"domain wall pinning", and "domain wall depinning region", respectively. The pinning region is the narrowest for the offset of 40% and becomes wider as the "$d$" value increases. Moreover, the pinning strength of the whole DW device increases significantly for the devices with $d = 60\%$, which is consistent with our R-H loop results. The depinning current density ($J_{dep}$) increases as offset increases. This is also shown in figure 3 (c) and table I for better understanding. From these results, we can see that pinning is weakest for $d = 40\%$ and strongest for $d = 60\%$.

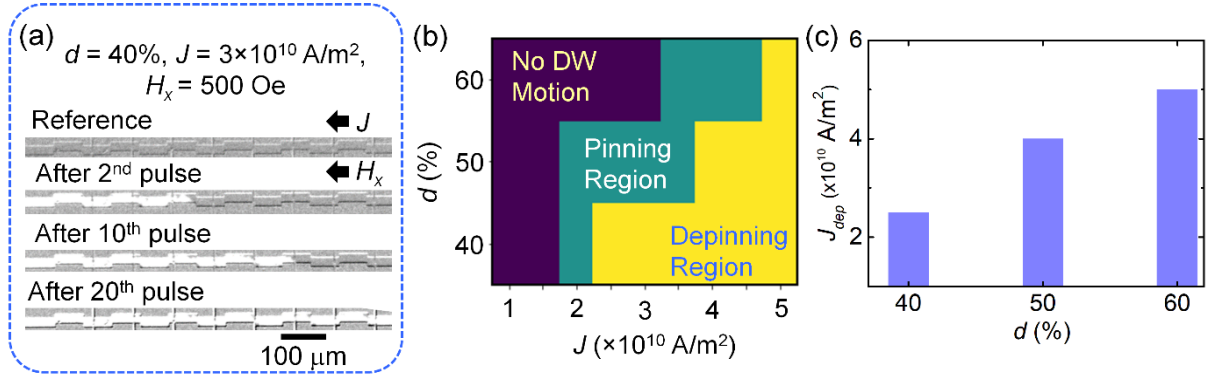

*Figure 3.* *(a) The results of Kerr microscopy to demonstrate the DW dynamics in meander synaptic devices with d = 40%, J = 3 × 10$^{10}$ A/m$^2$, and H$_x$ = 500 Oe. (b) The phase diagram of DW dynamics in meander devices as a function of "d" and "J". (c) The graph of depinning current density (J$_{dep}$) vs offset "d" values.*

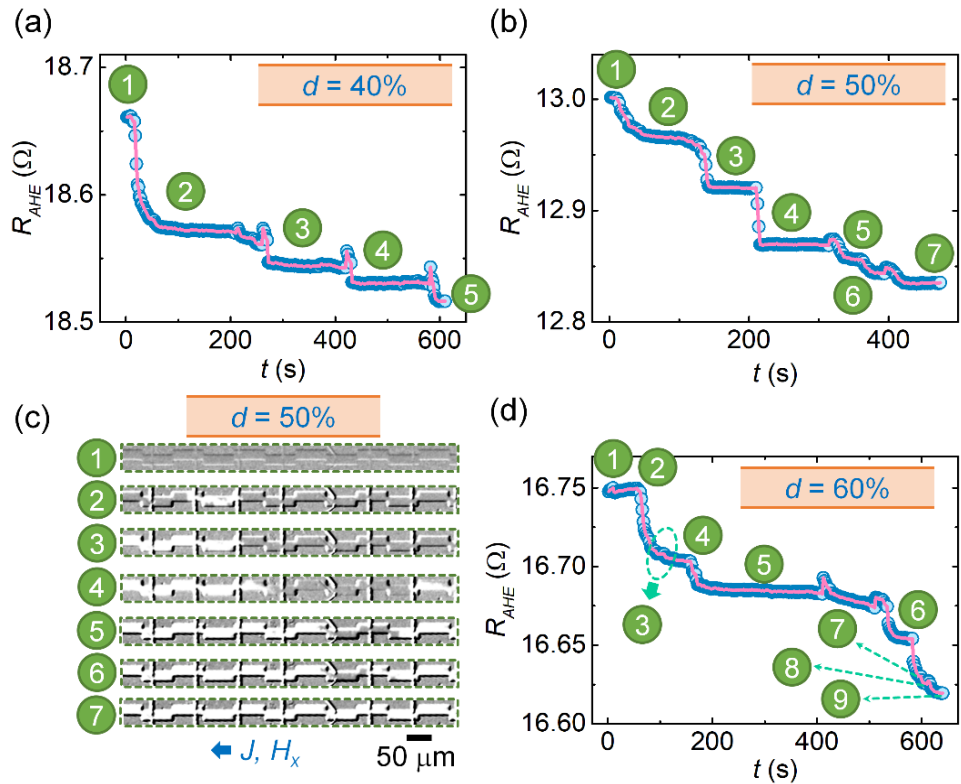

*Figure 4.* *The illustration of multiple resistance states for meander devices with an offset of (a) 40%, (b) 50%, and (d) 60%. (c) Kerr microscopy images, which show the systematic pinning and depinning at different pinning sites for an offset "d" of 50% (corresponding to figure 4 (b)).*

Furthermore, we performed experiments to study the multi-resistance states in meander devices (figure 4). The direction of the current density and in-plane magnetic field in these experiments is consistent with the previous experiments. Similarly, pulse width and time interval between the consecutive pulses remained the same. The magnitudes of the current densities were optimized to maximize the number of resistance states in all the devices. As observed in figure 4 (a), a total of 5 resistance states were obtained for $d = 40\%$. This number can be increased to 7 when $d$ was increased to 50% (figure 4 (b)). Figure 4 (c) shows the corresponding Kerr microscopy images, which illustrate the pinning of the DW at different pinning sites. The instantaneous magnetization state slightly differs when it is read using Kerr microscopy and the proposed method. This is because the Hall voltage reading using the proposed method is more sensitive to the change in magnetization in the vicinity of the read probes. When $d$ increases to 60%, the pinning becomes stronger, and the number of resistance states increases to 9. This result is presented in figure 4 (d). The summary of number of resistance states for all the studied meander devices is summarized in table I.

## Domain Wall-Based Neurons

### 1. Micromagnetic Simulations

For realizing the neuron functionalities, we studied the DW motion in straight DW devices using both simulations and experiments. First, we will discuss the results of the simulations. Similar to the synapse case, we first inserted ↓→↑ type DW and then applied the SOT current density ranging from $1 \times 10^{10}$ to $5 \times 10^{11}$ A/m$^2$. Under the influence of the SOT, the DW moves from the left end of the wire to the right end without any pinning. Depending on different configurations of reading elements, which will be discussed in the experimental section, we can realize the step and spike neuron functionalities. As one can see in figure 5 (a), the slope of the $m_z$ vs $t$ graph increases as the $J$ increases. This means that the DW velocity increases with $J$ (region I, figure 5 (a)). Here, $m_z/t$ is proportional to the DW velocity. However, beyond $J = 1 \times 10^{11}$ A/m$^2$, the slope becomes constant. This suggests that a maximum in SOT efficiency (for given magnetic parameters) occurs at $1 \times 10^{11}$ A/m$^2$ and it does not increase further (region III, figure 5 (b)). We have also observed a drop in domain wall velocity at $J = 7 \times 10^{10}$ A/m$^2$. This is because of the Walker breakdown during the DW motion in the perfect nanowire (region II, figure 5 (b)) [31-34]. Please refer to supplementary information 4 for a detailed discussion of these three regions of DW motion.

### 2. Experimental Results

As the first step in the experiments, we measured the R-H loops of the neuron devices. These results are presented in figure 5 (c). One can observe a sharp magnetization switching in both cases. This is an expected result for the case of the straight microwire as this does not offer any pinning (except the intrinsic pinning and/or extrinsic pinning from the edges of the device and extrinsic defects). The coercivity of the spike neuron device is 47 Oe, which is close to the thin film coercivity (figure 5 (d) and table II). However, the coercivity increases to 65 Oe for the step neuron device. This could be because of the following: The major difference between the spike and step neuron is the way reading probes were employed in these devices. For the case of spike neurons, we fabricated one pair of Hall crosses at the right end of the device. However, for step neurons, we fabricated eight pairs of equispaced Hall crosses spanning over the whole of the device. The Hall bar fabrication process induces some defects in the DW device. The number of such process-induced defects is more in the case of step neurons as compared to spike neurons. These can increase (*i*) the number of nucleation sites and (*ii*) the strength of the pinning sites. The latter may have a major role in increasing the coercivity of the step-neuron devices. Another interesting observation is that the difference in anomalous Hall resistance corresponding to up and down magnetization ($\Delta R_{AHE}$) has doubled for the case of the step neuron compared to the same in the spike neuron. This means that in addition to providing the non-local signal from the entire device and better SNR, our novel proposal also helps in increasing the sensitivity of the output signal. This is in line with our initial hypothesis. This aspect is also important as this may help in increasing the distinguishable multiple resistance states for the synaptic devices.

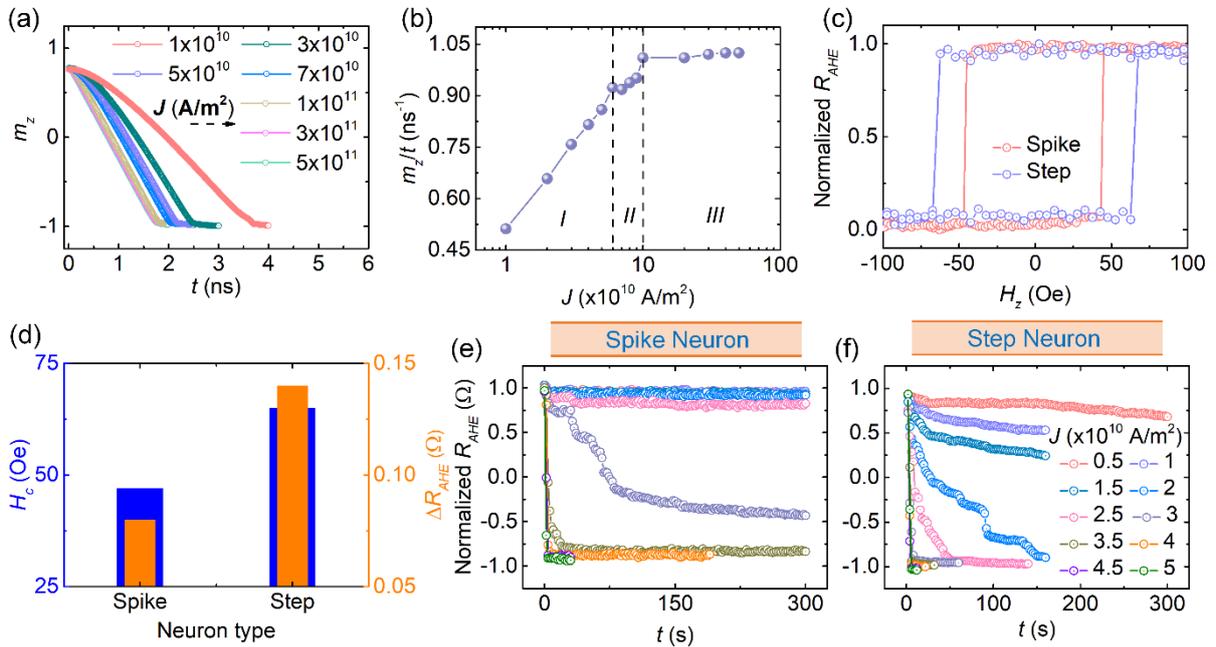

*Figure 5.* Simulation results: (a) The demonstration of domain wall (DW) motion in a straight (neuron) device for different current density values. (b) The plot of $m_z/t$ (∝ DW velocity) vs J for neuron devices. Experimental results: (c) The graph of $R_{AHE}$ vs OOP magnetic field for both types of neuron devices. (d) The comparison of coercivity and $\Delta R_{AHE}$ for two different types of neuron devices. SOT-driven DW motion for (e) spike and (f) step neuron devices for the current densities ranging from $5 \times 10^9$ A/m$^2$ to $5 \times 10^{10}$ A/m$^2$. An in-plane magnetic field of 500 Oe was applied in all cases.

Subsequently, we studied the SOT-driven domain wall motion in our neuron devices. The experimental procedure is the same as the one described for the synaptic device. The results of this set of experiments are presented in figures 5 (e-f) and table II. As one can see for spike neurons, the DW does not move (or negligible domain wall movement) for the current densities less than $2.5 \times 10^{10}$ A/m$^2$. This can be defined as region I of DW motion (table II). However, for the current density of $3.5 \times 10^{10}$ A/m$^2$ and above, the DW covers the whole DW devices in only a few current pulses (region III, in table II). The intermediate DW motion (region II) only occurs for the current density of $3 \times 10^{10}$ A/m$^2$. However, in the case of step neurons, the region

I of DW motion disappears. The slow and controlled DW motion region (region II), where several current pulses are required to move the DWs from one end to the other, expands from $J = 5 \times 10^9$ A/m$^2$ to $3 \times 10^{10}$ A/m$^2$. Similar to the case of spike neurons, for the current densities of $3.5 \times 10^{10}$ A/m$^2$ and above, fast domain wall motion was observed for the steps neuron devices as well. The reason for the above difference is the process-induced defects. When the electrical current pulses are applied, the nucleation of the reversed domains is easier for the step neurons. Once a domain wall is inserted in the domain wall devices, it can be moved in subsequent current pulses. However, for the spike neurons, the insertion of the DWs is harder due to fewer nucleation sites. We have also noticed that our HP3P3 thin film samples offer very less pinning site density in the as-deposited state [9]. The virtue of different DW motion nature because of process-induced defects also helps in achieving the distinguishable spike and step neuron functions.

*Table II: List of $H_c$ and $\Delta R_{AHE}$ and nature of DW motion for step and spike neuron devices.*

| Neuron Type | $H_c$ (Oe) | $\Delta R_{AHE}$ ($\Omega$) | Nature of Domain Wall Motion | | |
|---|---|---|---|---|---|
| | | | **No DW Motion (Region I)** | **Slow DW Motion (Region II)** | **Fast DW Motion (Region III)** |
| Spike | 47 | 0.08 | $5\times10^9$- $2.5\times10^{10}$ A/m$^2$ | $3\times10^{10}$ A/m$^2$ | $3.5\times10^{10}$ A/m$^2$ & above |
| Step | 65 | 0.14 | - | $5\times10^9$- $3\times10^{10}$ A/m$^2$ | $3.5\times10^{10}$ A/m$^2$ & above |

We then adjusted the applied current density or in-plane magnetic field or both to achieve the spiking and step neuron behavior in anomalous Hall resistance signal. For spike neuron devices, we first applied current pulses with a magnitude of $4.5 \times 10^{10}$ A/m$^2$ in the presence of an in-plane magnetic field (500 Oe). This results in the nucleation and expansion of a reversed domain. A magnetization state where all the spins point in the +z-axis was considered as the initial state (state 1 in figure 6 (a-b)). When we reverse the direction of the current, the magnetization direction changes. Since the read probe is only at the right end of the wire, a change in the magnetization occurs when the domain wall reaches the vicinity of the read probe (state 2 in figure 6 (a-b)). At this point, the direction of the current is reversed (same as state 1) to retain the initial magnetization state (state 3 in figure 6 (a-b)). We repeated this process to demonstrate two spikes in our experiments. The magnetization states 4 and 5 are identical to states 2 and 3 (or 1), respectively. We have also studied the domain wall dynamics in these spike neuron devices at different current densities. These results are presented in supplementary section 5. The magnitude and the direction of the in-plane magnetic remained the same during all these experiments.

Later, we performed the measurements for the step-neuron devices. First, we applied an electrical current density of magnitude $2 \times 10^{10}$ A/m$^2$ with a simultaneous in-plane magnetic field of 500 Oe. A reversed domain gets nucleated (state 1 in figure 6 (c-d)) and expands upon the application of the subsequent current pulses. States 2 and 3 represent intermediate magnetization states before we reverse the direction of the current. After 500 s, we reverse the direction of the current (backward direction). In the backward direction, $J = 3.5 \times 10^{10}$ A/m$^2$ and $H_x = 300$ Oe were utilized. The direction of the magnetic field remains unchanged. Here, the domain wall starts moving in the opposite direction. These are represented by states 4 and 5 of Figure 6 (c-d). Upon repeating the process of steps 1 through 5, identical results were obtained. The shift of initial state resistance to slightly higher values is attributed to the Joule heating of the devices.

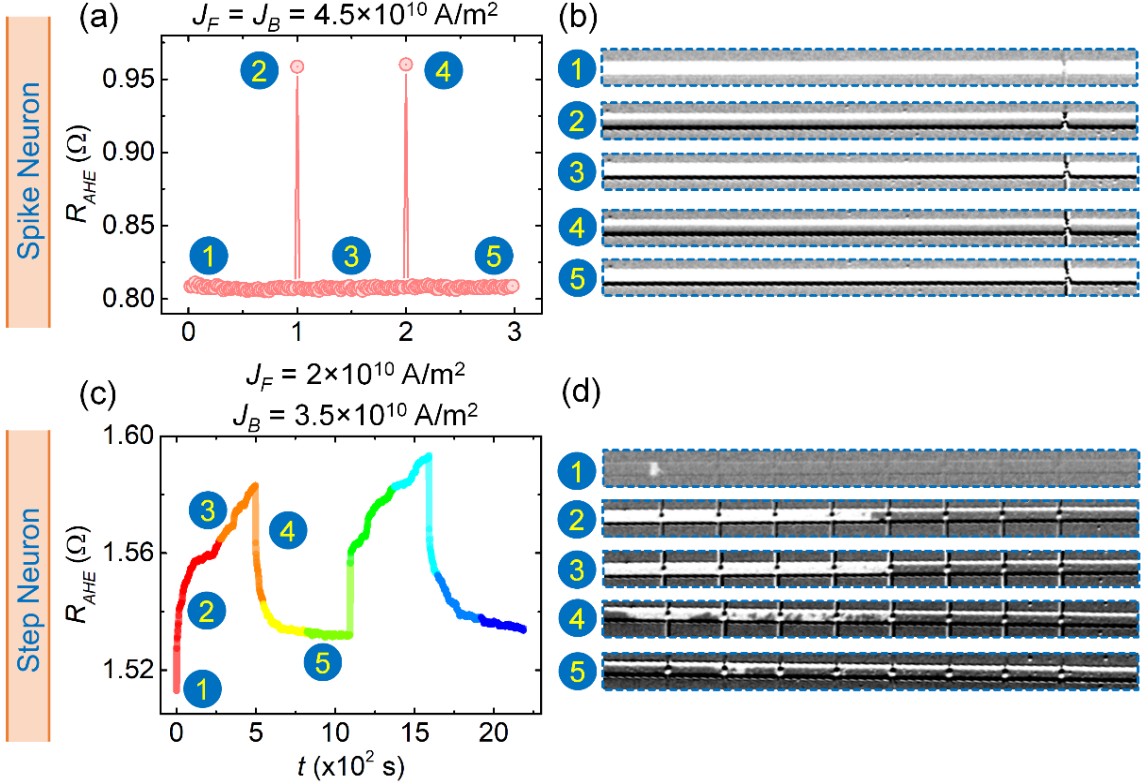

**Figure 6.** *SOT-driven domain wall (DW) motion for (a-b) spike and (c-d) step neuron devices. Figures (b and d) are Kerr microscopy images for different magnetization states (as indicated).*

## Conclusions

Owing to the suitability of DW devices for energy-efficient NC, we have studied DW device-based synapses and neurons through micromagnetic simulations and experiments. For synaptic operations, we proposed and studied the concept of meander DW devices. The depinning current density increases with the increase in offset. We found that an offset of ~50% is most suitable for synaptic devices. We studied a novel read probe that helped in improving the sensitivity and enhancing the SNR. We demonstrated nine distinct resistance states in our synaptic devices. For neurons, we studied straight DW devices in the form of spike and step neurons. For the spike neurons, we fabricated one pair of reading probes at the right end of the devices. However, step neurons utilize the same read strategy as in the meander devices. We found that the nature of the DW motion in step and spike neurons has a clear difference because of process-induced defects. This also helps in achieving the operations of step and spike functions and we successfully demonstrated the step and spike neuron functions in our devices. We believe our study contributes significantly toward the development of DW device-based NC architecture.

# Methods

**Micromagnetic Simulations:** Micromagnetic simulations were performed using Mumax$^3$ [27]. The dimensions of the meander devices with one pinning site and neuron devices were taken as 256 nm ($l$) × 32 nm ($w$) × 1 nm ($t$). Here, $l$, $w$, and $t$ stand for the length, width, and thickness of the simulated DW device. The length of the meander devices with 8 pinning sites increases to 1152 nm. The other two parameters remain the same. The cell size was taken as 1 nm × 1 nm × 1 nm. The magnetic parameters such as exchange constant ($A$), saturation magnetization ($M_s$), anisotropy constant ($K_u$), DMI constant ($D$), damping constant ($\alpha$), and spin Hall angle ($\theta_{SH}$) were chosen as 1.5×10$^{-11}$ J/m, 1×10$^6$ A/m, 1×10$^6$ J/m$^3$ (0, 0, 1), 0.5 mJ/m$^2$, 0.012, and 0.3 (except meander wires with 8 pinning sites, where $\theta_{SH}$ = 0.37), respectively. These parameters correspond to experimentally observed values in W/CoFeB/MgO system [9]. The following table provides the details of all the parameters, utilized during the simulations.

*Table 1: The list of parameters utilized during the micromagnetic simulations. These parameters can experimentally be realized for W/CoFeB/MgO material system [9].*

| S/No. | Parameter | Value |
|---|---|---|
| 1 | Device dimensions | 256 nm × 32 nm × 1 nm |
| 2 | Cell size | 1 nm × 1 nm × 1 nm |
| 3 | Exchange constant | 1.5 × 10$^{-11}$ J/m |
| 4 | Saturation magnetization | 1 × 10$^6$ A/m |
| 5 | Anisotropy constant | 1 × 10$^6$ J/m$^3$ |
| 6 | Easy-axis direction | (0, 0, 1) |
| 7 | Damping constant | 0.012 |
| 8 | DMI constant | 0.5 mJ/m$^2$ |
| 9 | Spin Hall angle | -0.3 |
| 10 | Exchange length | ~5 nm |
| 11 | Domain wall type | ↓→↑ |
| 12 | Current density | 1 × 10$^{10}$ A/m$^2$ - 1 × 10$^{12}$ A/m$^2$ |
| 13 | Meander offset | 10- 90 % |

**Thin Film Depositions:** The film stack Si/SiO$_2$/ HP-W (3 nm)/ LP-W (3 nm)/ Co$_{40}$Fe$_{40}$B$_{20}$ (1 nm)/ MgO (1 nm)/ Ru (2 nm) was deposited using DC/RF Singulus Timaris sputtering tool. Here, HP and LP mean the high- and low-pressure deposition. The HP-W films were deposited at an Ar gas pressure of 3.63 mTorr and a deposition power of 100 W (power density of 0.19 W/cm$^2$). The same for LP-W was 0.78 mTorr and 100 W. Such a low power density results in a deposition rate of 0.03 nm/s, which is essential for $\beta$- phase in W thin films. After depositing the thin film samples, we characterized them using polar Kerr microscopy and measured the magnetic hysteresis (M-H) loop. The SOT properties were characterized using resistivity measurements and second Harmonic measurements. Please refer to reference [9] for all the details of thin film properties.

**Device Fabrication and Measurements:** We fabricated micrometer-sized DW-based synaptic and neuron devices using optical lithography and ion milling. The dimensions of these devices were 900 μm ($l$) × 20 μm ($w$) × 1 ($t$). Note, that the stack thickness is 10 nm. The offsets were chosen to be 40, 50, and 60%. For neuron devices, we fabricated a geometrical pinning site in the form of a meander structure to stop the DW at the right end and revert. Subsequently, we fabricated Hall crosses using optical lithography and a lift-off process. A 30 nm thick Ta layer was utilized to form the Hall crosses. The width of the Hall crosses was chosen to be 5 μm. After this, we fabricated electrode pads of dimension 300 μm × 300 μm using optical lithography and a lift-off process. The electrode material consists of Ta (35-40 nm)/ Cu (90 nm)/ Ta (10 nm).

**Electrical Characterizations:** Once the devices were fabricated, we used Kerr microscopy system set up with a vector magnet, Keithly 6221 current source, and Keithly 2182A nano-voltmeter to perform the measurements.

To measure the change in magnetization while sweeping the out-of-plane (OOP) magnetic field, we measured the R-H loops. For this, we first saturated the devices with a large OOP magnetic field of -1000 Oe. Then we swept the OOP magnetic field from -ve to +ve and then +ve to -ve field directions. The corresponding anomalous Hall voltage was measured at every field value. For this, we applied the pulsed reading current of 100 μA with a pulse width of 0.2 s. A time interval of 2 s was set between consecutive field values. From the voltage values, anomalous Hall resistance was calculated using Ohm's relation.

For current-induced DW motion experiments, we again saturated the devices with an OOP magnetic field of -1000 Oe. Then, we applied the current pulses of a certain amplitude with a pulse width of 0.2 s. The time interval between subsequent pulses was fixed at 2 s. A simultaneous longitudinal magnetic field of 500 Oe was applied in all cases. The current density was varied from $5 \times 10^9$ A/m$^2$ to $5 \times 10^{10}$ A/m$^2$ in most of the cases (unless specified). Similar to the previous cases, the output was read using the anomalous Hall effect.

## Acknowledgments

The authors gratefully acknowledge the National Research Foundation (NRF), Singapore for the NRF-CRP (NRF-CRP21-2018-0003) grant. The authors also acknowledge the support provided by Agency for Science, Technology and Research, A*STAR RIE2020 AME Grant No. A18A6b0057 for this work.

## Conflict of Interest

The authors declare no conflict of interest.